
\overfullrule=0pt
\magnification=\magstep1
\baselineskip= 1.4\baselineskip
\raggedbottom
\font\fivepoint=cmr5
\headline={\hfill{\fivepoint JKFEHL -- 29/June/94}}
\def\R{{\bf R}}
\def\d{{\rm d}}
\def\q{{\bf q}}
\def\p{{\bf p}}

\def\H{{\cal H}}
\def\e{{\bf e}}
\def\E{{\cal E}}
\def\C{{\bf C}}
\def\B{{\cal B}}
\def\an{{\rm an}}
\def\Tr{{\rm Tr}}

\def\1{{\bf 1}}

\def\uprho{\raise1pt\hbox{$\rho$}}
\def\mfr#1/#2{\hbox{${{#1} \over {#2}}$}}
\font\eightpoint=cmr8
\def\upvarphi{\raise1pt\hbox{$\varphi$}}
\def\upchi{\raise1pt\hbox{$\chi$}}
\centerline{\bf THE GROUND STATE OF A GENERAL ELECTRON-PHONON}
\centerline{\bf HAMILTONIAN IS A SPIN SINGLET}
\vskip 1truecm
{\baselineskip=3ex
\centerline{J. K. Freericks$^{(1),}$\footnote{$^{\dagger}$}{\baselineskip=12pt
%
%
\eightpoint
Address after Sept. 1, 1994:  Department of Physics,
Georgetown University, Washington, DC
20057-0995}
and Elliott H. Lieb$^{(2)}$
%
%
\vfootnote{}{\eightpoint \copyright 1994 by the authors.
Reproduction of this article by any means is permitted for non-commercial
purposes.}}}
\bigskip
\noindent{\it $^{(1)}$Department of Physics, University of California,
Davis, CA  95616 \hfil\break
$^{(2)}$Departments of Mathematics and Physics,
Princeton University, P.O. Box 708, Princeton, NJ, 08544-0708\hfil\break
}
\bigskip
\bigskip
{\narrower{\it Abstract:\/} The many-body ground state of a very
general class of
electron-phonon Hamiltonians is proven to contain a spin singlet
(for an even number of electrons on a finite lattice).  The phonons interact
with the electronic system in two different ways---there is an interaction with
the local electronic charge and there is a functional
dependence of the electronic hopping Hamiltonian on the phonon coordinates.
The phonon potential energy may include anharmonic terms, and the
electron-phonon couplings and the hopping matrix elements may be
nonlinear functions of the phonon coordinates.  If the hopping Hamiltonian
is assumed to have no phonon coordinate dependence, then
the ground state is also shown to be unique, implying that there are no
ground-state level crossings, and that the ground-state energy is an analytic
function of the parameters in the Hamiltonian.  In particular, in a finite
system any self-trapping
transition is a smooth crossover not accompanied by
a nonanalytical change in the ground state.
The spin-singlet theorem applies to the Su-Schrieffer-Heeger model and
both the spin-singlet and uniqueness theorems apply to the
Holstein and attractive Hubbard models as special cases.  These results
hold in all dimensions --- even on a general graph without periodic lattice
structure.\smallskip}
\bigskip
\bigskip
\bigskip
\bigskip
\bigskip
\bigskip

\centerline{\bf I. INTRODUCTION}
\bigskip
Electrons have a tendency to pair when the effective electron-electron
interaction has an attractive region; in particular this occurs when
electrons interact by exchanging bosons.  The resulting ground state then often
sustains either
superconducting or charge-density-wave order.  The simplest interacting
Hamiltonian of this type
is one in which electrons interact indirectly with each other
via phonons.  Migdal$^1$ analyzed the electron-phonon
interaction in the normal state and discovered that in the limit in which the
phonon frequency $\Omega$ is much smaller than the Fermi energy $E_f$, the
full many-body
theory can be described by a first-order, self-consistent Hartree-Fock theory,
and that the neglected higher-order diagrams (vertex corrections) usually
contribute
to order $\Omega/E_f$.  This result has been named Migdal's theorem and it
classifies those nonadiabatic processes that are typically important
for describing
low-frequency electron-phonon interactions.  Soon thereafter, Eliashberg$^2$
generalized Migdal's result to the superconducting phase and discovered that
a similar first-order self-consistent Hartree-Fock theory would describe
superconductivity.  Rowell and McMillan$^3$ subsequently demonstrated that
one
could directly measure the electron-phonon spectral density from tunneling
experiments and then use the formalism of Migdal and Eliashberg to describe
all of the remaining properties of the superconducting state.
Migdal-Eliashberg theory has been successful in predicting
transition temperatures (and other materials properties) of most low
temperature superconductors$^{4,5}$.

The electron-phonon Hamiltonian considered here is
$$H=\sum_{\sigma}\sum_{x, y \in \Lambda} t_{x y}({\bf q})
c_{x\sigma}^{\dag}
c^{{\phantom{\dagger}}}_{y \sigma}+\sum_{x\in \Lambda}G_x({\bf q})
(n_{x\uparrow}+
n_{x\downarrow})+{1\over 2}\sum_{j=1}^\nu (M_j \Omega_j^2 q_j^2 +
{1\over M_j} p_j^2)+V_{\an}({\bf q}) .\eqno(1.1)$$
We can also add to this an attractive Hubbard type interaction --- as
discussed later in Sect.~III.
Our notation is the following:  The electrons occupy positions on a finite
``lattice'' or ``graph'' $\Lambda$, which is some collection of $|\Lambda|$
sites;
we emphasize that no specific
periodicity or dimensionality is assumed.  The operator
$c_{x\sigma}^{\dag}$ $(c^{{\phantom{\dagger}}}_{x\sigma})$ is a creation
(annihilation) operator for an electron at lattice site $x$ with $z$-component
of spin $\sigma = \uparrow$ or $\downarrow$.  These operators satisfy the
anticommutation relations $c^\dagger_{x \sigma} c^{{\phantom{\dagger}}}_{y
\sigma} + c^{{\phantom{\dagger}}}_{y \sigma}
c^\dagger_{x \sigma} = \delta^{{\phantom{\dagger}}}_{xy}$ and
$c^{{\phantom{\dagger}}}_{x \sigma} c^{{\phantom{\dagger}}}_{y \sigma} +
c^{{\phantom{\dagger}}}_{y
\sigma} c^{{\phantom{\dagger}}}_{x \sigma} = 0$ for each
$\sigma = \uparrow$ or $\downarrow$.  It
is customary to assume that the up-spin operators also anticommute with the
down-spin operators, but it is more convenient for us to assume that they
commute, i.e., $c_{x \uparrow} c_{y \downarrow} - c_{y \downarrow} c_{x
\uparrow} = 0$, etc..  This change is innocuous [as long as particle number
is conserved --- which it is with the Hamiltonian (1.1)] and is effected
by replacing $c_{x \uparrow}$ by the operator $\exp [i \pi N_\downarrow]
c_{x \uparrow}$ with $N_\sigma := \Sigma_{x \in \Lambda} n_{x \sigma}$ and
$n_{x \sigma} := c^\dagger_{x \sigma} c_{x \sigma}$; the operator
$c^\dagger_{x \uparrow}$ is replaced similarly, but
$c^{{\phantom{\dagger}}}_{x \downarrow}$ and
$c^\dagger_{x \downarrow}$ are unchanged.

The phonon modes are indexed by $j$ and, for technical simplicity, we
assume there are finitely many of them, namely $\nu$.  In some special
models, such as the Holstein model, defined in (1.9),
there is an association between the
phonon modes and the lattice sites, but such an association is neither
required nor assumed.  The phonon coordinates are $q_1, \dots , q_\nu$
and the momenta are $p_1, \dots , p_\nu$, denoted collectively by $\q$
and $\p$.

The $p_j$'s and $q_j$'s satisfy the usual commutation relations $[q_j, p_k]
= i \delta_{j k}$ and we shall represent these in the usual way as operators
on $L^2 (\R^\nu)$, the set of square integrable functions of $\nu$
variables, by $p_j = - i \d/\d q_j$ (with $\hbar = 1$).

The most general positive-definite quadratic form can always be put in the
normal mode form, shown in (1.1), in which the numbers $M_j > 0$ and
$\Omega_j > 0$ are, respectively, the masses and frequencies of the
corresponding phonon normal modes.  For convenience, we explicitly exclude
zero frequency modes, which, physically, correspond to center of mass
translation.  The additional potential term $V_{an} (\q)$ includes all
non-quadratic terms; it is completely arbitrary except for the assumption
that it is bounded below, i.e., $V_{\an} (\q) \geq C$ for some number $C$,
and that $\sum \nolimits^\nu_{j=1} M_j \Omega^2_j q^2_j + V_{\an} (\q)$ goes to
infinity faster than linearly in all directions.

The real hopping matrix $t(\q)$, whose elements are $t_{xy} (\q)$, is
allowed to be an arbitrary measurable function of the phonon coordinates
(but {\it not} the momenta).  An important assumption is that $t(\q)$ is real
and symmetric for each $\q$, i.e., $t_{xy} (\q) = t_{yx} (\q)$.  We also
assume,
for convenience, that $\Tr \,\vert t(\q) \vert$ is finite for every $\q$.
[Here, $\vert t (q) \vert = \sqrt{t(\q)^2}$.]  We do not make any
assumption about the relative signs and magnitudes of the hopping matrix
elements.  The reality assumption generically precludes the interaction of
the electronic orbital motion with magnetic fields.

The electron-phonon coupling $G_x({\bf q})$ is
also an arbitrary real function of the phonon coordinates
that couples the phonons to the total electronic
charge at lattice site $x$.  For Theorem 1 (existence of singlet ground
states) the only assumption about these couplings is
lower boundedness of the {\it total} phonon potential energy, i.e., we
assume that the function of $\q$ given by
$$-\Tr \,\vert t(\q)\vert -2 \sum \limits_{x \in \Lambda} \vert G_x (\q)
\vert + {1 \over 2} \sum \limits^\nu_{j=1} M_j \Omega_j^2 q^2_j +
V_{\an} (\q)$$
is bounded below.  Usually, one assumes that $G_x (\q)$ is a linear
function of $\q$, but we do not do so.

The total spin is a conserved quantity of the Hamiltonian $H$ in Eq. (1.1). The
spin operators are defined to be the quadratic operators
$$S^z:= {1\over 2}\sum_{x\in \Lambda}(n_{x\uparrow}-
n_{x\downarrow}),\quad S^+:= (S^-)^{\dag}:=
\sum_{x\in \Lambda} c^{\dag}_{x\uparrow}
c^{{\phantom{\dag}}}_{x\downarrow}. \eqno(1.3)$$
They all commute with the Hamiltonian $H$.  The spin operators satisfy an
SU(2) algebra, and the total-spin operator is defined to be the corresponding
quadratic Casimir operator $(S_{\rm op})^2:= (S^z)^2 + {1\over 2} S^+
S^-+{1\over 2}
S^-S^+$, with eigenvalues $S(S+1)$.  In particular, we are interested in the
eigenvalues of the total-spin operator for the ground states of $H$
of the electron-phonon model described in Eq. (1).

Our main result asserts that the ground state of the model in (1.1) has an
$S = 0$ ground state, and that the ground state is often unique.

{\bf THEOREM 1 (Existence of singlet ground states).}  {\it  Assume the
previously stated conditions on the Hamiltonian $H$ in (1.1).  Assume
additionally, that the total number of electrons, $2N$, is even. Then among all
of the ground states of $H$ there is at least one ground state with total spin
$S=0$.}

For Theorem 2 (uniqueness of the
ground state) additional assumptions are needed.

(i) The hopping matrix elements $t_{xy}$ are independent of ${\bf q}$.  We also
assume that $\Lambda$ is {\bf connected}, i.e., for each $x$ and $y$ in
$\Lambda$ there are sites $x = x_0, x_1, x_2, \dots , x_n = y$ such that
$t_{x_i x_{i+1}} \not= 0$ for all $0 \leq i \leq n-1$.  A {\bf bond} is
said to exist between two sites $x$ and $y$ in $\Lambda$ if $x \not= y$ and
if $t_{xy} \not= 0$.

(ii) All the functions of ${\bf q}$ appearing in $H$, i.e.,
$G_x ({\bf q}), V_{\an} ({\bf q})$,
are differentiable.  [Actually, it suffices for them to be locally H\"older
continuous with densely defined derivatives.]

(iii) The $G_x ({\bf q})$'s are {\bf independent}.
By this we mean that for each
point ${\bf q} \in \R^\nu$ the $\nu$ simultaneous equations
$$\sum \limits_{x \in \Lambda} {\partial G_x ({\bf q})
\over \partial q_j} A_x =
0, \qquad j = 1,2, \dots , \nu \eqno(1.2)$$
have no common solution other than $A_x = 0$ for all $x \in \Lambda$.  In
other words the $(\vert \Lambda \vert \times \nu)$ matrix $\partial G_x
/\partial q_j$ has rank $\vert \Lambda \vert$ for each $\q \in \R^\nu$.
(Again, it suffices for this to hold only on a dense subset of $\R^\nu$.)

(iv)  Every mass, $M_j$, is finite.

These conditions hold in many models, e.g., the Holstein model$^6$, but
conditions (i) and (iii) do not hold in the Su-Schrieffer-Heeger (SSH)
model$^7$.

{\bf THEOREM 2 (Uniqueness of the ground state).}  {\it
If, (i)-(iv) above are satisfied
then the ground state is unique (and hence a nondegenerate spin singlet).}

{\it Remarks.}---(1) Theorem 1 has long been conjectured and is consistent
with the intuition that
the exchange of a boson leads to electron-electron pairing.

(2) The uniqueness theorem establishes that the many-body ground state of $H$
does not have any level crossings for a finite system, thereby
establishing the result
that the self-trapping transition from a collection of extended polarons
to a collection of localized polarons is a smooth crossover, rather than
a transition by breaking of analyticity in any finite system.

The proof of these theorems is based upon spin-reflection positivity and is
closely related to the analogous proof for the Hubbard model already presented
by one of us$^8$.  A different proof based upon Perron-Frobenius
positivity arguments, was given for one-dimensional models$^9$, but
it does not appear
to be readily generalizable to the present case.

The SSH model is the special case of (1.1) in which
the hopping matrix elements are {\it linear} functions of the phonon
coordinates
and the electron-phonon couplings vanish [$G_x({\bf q})=0$].
To be more precise consider the
original SSH model on a periodic one-dimensional chain$^7$
$$H_{SSH}=\sum_{\sigma}\sum_{i=1}^{|\Lambda|}(t-\delta t [Q_{i+1}-Q_i])(
c_{i+1\sigma}^{\dag} c^{{\phantom{\dagger}}}_{i\sigma} + c_{i\sigma}^{\dag}
c^{{\phantom{\dagger}}}_{i+1\sigma}) +
{1\over 2}\sum_{i=1}^{|\Lambda|}[\kappa (Q_{i+1}-Q_i)^2+{1\over M}P_i^2]
\quad ,\eqno(1.4)$$
with $Q_i$ $(P_i)$ the local phonon coordinate (momentum) at site $i$.
Transforming to the normal coordinates
$$q_j :={1\over \sqrt{\vert \Lambda \vert}} \sum_{k=1}^{|\Lambda|}Q_j
\left[ {\cos 2\pi k
{j - {1\over 2}|\Lambda|\over |\Lambda |} \quad\quad\quad {1\over 2}|\Lambda|
\le j < |\Lambda|\atop
\sin2\pi k {j -{1\over 2}|\Lambda|\over|\Lambda|}\quad\quad\quad 0\le j <
{1\over 2}|\Lambda|}\right.\eqno(1.5)$$
yields the electron-phonon Hamiltonian in the form of Eq. (1.1)
$$H_{SSH}=\sum_{\sigma}\sum_{i=1}^{|\Lambda|}[t-T_{i+1}({\bf q})](
c_{i+1\sigma}^{\dag} c^{{\phantom{\dagger}}}_{i\sigma} + c_{i\sigma}^{\dag}
c^{{\phantom{\dagger}}}_{i+1\sigma})+
{1\over 2} \sum_{j=0}^{|\Lambda|-1} [M\Omega_j^2 q_j^2 + {1\over M}
p_j^2]\quad , \eqno(1.6)$$
with
$$T_i({\bf q})=\sum_{j=0}^{|\Lambda|/2-1} (-1)^i q_j \sin{2\pi ij
\over|\Lambda|}+
\sum_{j= |\Lambda|/2}^{|\Lambda|}(-1)^i q_j \cos{2\pi ij \over|\Lambda|}
\quad ,\eqno(1.7)$$
and $\Omega_j^2 = 2\kappa[1+\cos(2\pi j/|\Lambda|)]/M$.
Theorem 1 shows that the SSH model always contains
a spin-singlet ground state,
but the ground state is not necessarily unique.  We are aware of no other
rigorous results for the SSH model.

The Holstein model$^6$ is a special case of (1.1) where there is one
(internal) normal mode associated with each
lattice site (the index $j$ is identical to the index $x$), the
hopping matrix elements have no phonon coordinate dependence, the
electron-phonon coupling is linear in the phonon coordinate associated with
the lattice site
$$G_x({\bf q})=g_xq_x\quad ,\eqno(1.8)$$
and the potential energy is harmonic $[V_{an}({\bf q})=0]$.  The resulting
Holstein Hamiltonian is
$$H_{Hol}=\sum_{\sigma}\sum_{x, y \in \Lambda} t_{xy} c_{x\sigma}^{\dag}
c^{{\phantom{\dagger}}}_{y \sigma}+\sum_{x\in \Lambda}g_xq_x(n_{x\uparrow}+
n_{x\downarrow})+{1\over 2}\sum_{x\in\Lambda}(M_x\Omega_x^2q_x^2+
{1\over M_x}p_x^2)\quad .\eqno(1.9)$$
The Holstein model has independent couplings if all $g_x$ are nonzero.
The only rigorous result for the Holstein model is that of L\" owen
$^{10}$ for {\it one} electron. In this case, it has been shown that the
ground state is nondegenerate and
analytic if the lattice is bipartite.

In the static limit, where all of the phonon masses become infinite, but the
spring constant remains finite
$$M_x\rightarrow \infty\quad ,\quad\kappa_x=M_x\Omega_x^2={\rm finite}\quad ,
\eqno(1.10)$$
the phonon kinetic energy terms $\sum_xp_x^2/2M_x$ do not contribute to the
Hamiltonian (1.9).  The up- and down-spin electrons become independent and
the Holstein model maps onto a Falicov-Kimball model$^{11}$
with spin one-half conduction electrons and a {\it continuous}
static field $q_x$.
Techniques similar to those used in the spinless Falicov-Kimball model$^{12}$
may be used to show that the ground state is a commensurate charge-density-wave
at half filling$^{13,14}$.  The static limit of the Holstein model has also
been investigated by other methods$^{15}$ and shown to possess insulating
bipolaronic charge-density-wave order at large enough coupling.  The ground
state has also been shown to be nonanalytic in one dimension$^{15}$.

In the instantaneous limit, where the phonon frequency and electron-phonon
coupling become infinite, but their ratio remains finite,
$$g_x\rightarrow\infty\quad ,\quad\Omega_x\rightarrow\infty\quad ,\quad
{g_x\over \Omega_x}={\rm finite}\quad ,\eqno(1.11)$$
the Holstein model maps onto an attractive Hubbard model$^{16}$.  This mapping
is illustrated by completing the square in Eq. (1.9)
$$H_{Hol}=\sum_{\sigma}\sum_{x,y \in \Lambda} t_{xy} c_{x\sigma}^{\dag}
c^{{\phantom{\dagger}}}_{y \sigma} - {1\over 2}\sum_{x\in \Lambda}
U_x(n_{x\uparrow}+ n_{x\downarrow})^2$$
$$\quad +{1\over 2}\sum_{x\in\Lambda}
(M_x\Omega_x^2[q_x +
{g_x\over M_x\Omega_x^2}\{n_{x\uparrow}+n_{x\downarrow}\}]^2+
{1\over M_x}p_x^2)\quad ,\eqno(1.12)$$
with the electron-electron interaction $U_x$ defined to be
$$U_x:= {g_x^2\over M_x\Omega_x^2}\quad .\eqno(1.13)$$
In this instantaneous limit
the remaining electron and phonon terms in the Hamiltonian decouple because
$g_x/M_x \Omega^2_x \rightarrow 0$, and
one is left with an attractive Hubbard model
$$H_{Hub}=\sum_{\sigma}\sum_{x,y \in \Lambda}\bar
t_{xy} c_{x\sigma}^{\dag}
c^{{\phantom{\dagger}}}_{y \sigma} - \sum_{x\in \Lambda} U_xn_{x\uparrow}
n_{x\downarrow} +{1\over 2}\sum_{x\in\Lambda}(M_x\Omega_x^2q_x^2+
{1\over M_x}p_x^2)\quad ,\eqno(1.14)$$
with
$$\bar t_{xy} := t_{xy} - {1\over 2}U_x\delta_{xy} \quad .\eqno(1.15)$$
The attractive Hubbard model is already known to have a unique spin-singlet
($S=0$) ground state for an even number of electrons on a finite lattice$^8$.

In Section II, the proofs of Theorems 1 and 2 are presented
for the Hubbard model in order to clarify the results of Ref.~8 and to
define our notation and methodology.  Section III contains the proofs of
these theorems for the electron-phonon Hamiltonian.
A discussion of the results follows in Section IV.

\bigskip
\centerline{\bf II. ATTRACTIVE HUBBARD MODEL PROOFS}
\bigskip

Proofs are presented for the results in Ref.~8 in order to clarify the
previous work and to define the notation and current methodology.
We begin with a proof of Theorem 1.  The attractive Hubbard model
Hamiltonian is given by the electronic terms in Eq. (1.14) with each
$U_x\ge 0$ and the bar dropped from the hopping matrix (the hopping matrix
elements have no phonon coordinate dependence here).

{\bf Proof of Theorem 1 for the attractive Hubbard model.}---Both
the total spin operator $S^2$ and the $z$-component of spin $S^z$ are
conserved quantities of the Hubbard model (1.14) and restriction can be made
to the $S^z=0$ subspace (without loss of generality), because every eigenstate
with total-spin $S$ can be rotated into the subspace with $S^z=0$ without
changing its energy.  Therefore, we can assume there are $N$ electrons of each
spin (up and down).

It is convenient to use first quantized notation.  We denote the
coordinates of the up-spin electrons with $X$ which really is an $N$-tuple
$X = (x_1, x_2, \dots , x_N)$ where each $x_i \in \Lambda$.  Similarly $Y$
denotes the coordinates of the down-spin electrons.  Any wave function
$\Psi$ is a function of both coordinates $\Psi (X,Y)$ and it is
antisymmetric in $\{ x_i \}^N_{i=1}$ and antisymmetric in $\{ y_i
\}^N_{i=1}$.  Note that if $\Psi (X,Y)$ is an eigenfunction, then so is
$\Psi (Y,X)$ and $\Psi (Y,X)^*$.  (It is here that the condition that the
hopping matrix elements $t_{xy}$ are real is used.)  Instead of considering
an eigenfunction $\Psi (X,Y)$ it is convenient to consider $\Psi (X,Y) +
\Psi (Y,X)^*$ and $i [ \Psi (X,Y) - \Psi (Y,X)^*]$.  In other words we can,
without loss of generality, assume that our eigenfunction, viewed as a
matrix indexed by $X$ and $Y$, is self adjoint (but not necessarily real).
$$\Psi (X,Y) = \Psi (Y,X)^*. \eqno(2.1)$$
The dimension, $d$, of this matrix $\Psi$ is
$$d = {\vert \Lambda \vert \choose N}. \eqno(2.2)$$

Any self-adjoint matrix can be expanded in terms of its eigenfunctions.
Thus,
$$\Psi (X,Y) = \sum \limits^d_{\alpha =1} w_\alpha \phi_\alpha (X)
\phi_\alpha (Y)^* \eqno(2.3)$$
where the $\phi_\alpha$'s are an orthonormal set of functions (but
antisymmetric in their argument $X$), and the $w_\alpha$'s are real numbers.
Our aim is to show that the $w_\alpha$'s can all be chosen to be nonnegative.
This will conclude the proof because it implies that
$$\sum \limits_X \psi (X,X) > 0, \eqno(2.4)$$
which implies that $\psi (X_0, X_0)$ is positive for at least one $X_0$.
This means that the wave function does not vanish when the up-spin
electrons and the down-spin electrons are at precisely the same locations
--- and this is a singlet state.  Thus $\psi$ has a nonvanishing component
in the $S= 0$ sector.

To show that the $w_\alpha$'s can be taken nonnegative, let us write out
the energy using the decomposition of $\Psi$ in (2.3).  One easily computes
$$\eqalignno{\langle \Psi \vert H \vert \Psi \rangle &= 2 \sum
\limits^d_{\alpha =1} w^2_\alpha \langle \phi_\alpha \vert K \vert
\phi_\alpha \rangle
- \sum \limits^{\vert \Lambda \vert}_{x=1} U_x \sum \limits^d_{\alpha = 1}
\sum \limits^d_{\beta = 1} w_\alpha w_\beta \big\vert \langle \phi_\alpha
\vert L_x \vert \phi_\beta \rangle \big\vert^2 \qquad&(2.5)\cr
\langle \Psi \vert \Psi \rangle &= \sum \limits^d_{\alpha = 1} w^2_\alpha
\langle \phi_\alpha \vert \phi_\alpha \rangle \qquad&(2.6)\cr}$$
where the $d$-dimensional matrices $K$ and $L_x$ are defined as follows:
$K$ is the first-quantized version of $\sum \nolimits_{xy} t_{xy}
c^\dagger_x c^{{\phantom{\dagger}}}_y$ (no spin here) and $L_x$ is the
first-quantized version of $n_x$.  More explicitly, the matrix elements
appearing in (2.5) and (2.6) are constructed in the following manner:  The
inner product between two arbitrary vectors $\phi_1 (X)$ and $\phi_2 (X)$
is
$$\langle \phi_2 \vert \phi_1 \rangle = \sum \limits_X \phi_2 (X)^* \phi_1
(X). \eqno(2.7)$$
The kinetic energy matrix elements satisfy
$$\langle \phi_2 \vert K \vert \phi_1 \rangle = \sum \limits_{xy} t_{xy}
\sum \limits^N_{j=1} \sum \limits_X \phi_2 (x_1, \dots , x, \dots , x_N)^*
\phi_1 (x_1, \dots , y, \dots , x_N) \eqno(2.8)$$
where the argument of $\phi_2$ agrees with the argument of $\phi_1$
everywhere except at the $j^{th}$ index where the site index is $x$ for
$\phi_2$ and $y$ for $\phi_1$.  The number operator matrix elements
satisfy
$$\langle \phi_2 \vert L_x \vert \phi_1 \rangle = \sum \limits_X \phi_2
(X)^* \phi_1 (X) \sum \limits^N_{j=1} \delta_{x_j, x} \eqno(2.9)$$

For the purpose of Theorem 1, the explicit values of these matrix elements
are unimportant.  The only thing one has to note about (2.5) and (2.6) is
that replacing every $w_\alpha$ by $\vert w_\alpha \vert$ cannot increase
the energy.  The first term in (2.5) and the inner product of (2.6) stay
the same, while the second term in (2.5) can only improve (if it changes at
all).  Thus if $\Psi$, given by (2.3) is a ground state of the Hubbard model
then so is $\vert \Psi \vert$ which is constructed by replacing $w_\alpha
\rightarrow \vert w_\alpha \vert$ in (2.3).  Note that $\vert \Psi \vert
(X,Y)$ is not equal to $\vert \Psi (X,Y) \vert$ in general, but corresponds
to $\vert \Psi \vert = \sqrt{\Psi^2}$ in the sense of matrices.  {\bf Q.E.D.}

In order to prove Theorem 2 we first need to establish a lemma and in
order to state the lemma a definition is needed.  If $X = (x_1, \dots ,
x_N)$ we define the operator $\Pi^X$ to be
$$\Pi^X := L_{x_1} L_{x_2} \dots L_{x_N} \eqno(2.10)$$
Since the different $L_x$'s commute, the ordering of factors in (2.10) is
unimportant.  In second quantized notation $\Pi^X = n_{x_1} \dots n_{x_N}$
which shows that $\Pi^X$ is an orthogonal projector, i.e., $\Pi^X =
(\Pi^X)^\dagger$ and $(\Pi^X)^2 = \Pi^X$.  It is also obvious that $\Pi^X$
is a {\it one-dimensional} projector.  Furthermore, if $X$ and $X^\prime$
differ only by a permutation then $\Pi^X = \Pi^{X^\prime}$.  The matrix
elements of $\Pi^X$ [analogous to (2.9)] satisfy
$$\langle \phi_2 \vert \Pi^X \vert \phi_1 \rangle = \phi_2 (X)^* \phi_1 (X)
N!. \eqno(2.11)$$
[The $N!$ in (2.11) may appear mysterious, but it is not.  The reason
is that if $\phi_1$ and $\phi_2$ are normalized vectors concentrated at $X$ and
all of its $N!$ permutations and at no other $X$, then $\vert \phi_1 (X)
\vert^2 = \vert \phi_2 (X) \vert^2 = 1/N!$.  Thus, both
sides of Eq.(2.11) are equal in magnitude to 1.]

{\bf LEMMA (Connectivity of the single-spin configuration space).}  {\it
Assume that the lattice $\Lambda$ is connected as explained above.  Then
the single-spin configuration space is connected by the kinetic energy
matrix $K$.  That is to say, given points $X = (x_1, \dots ,x_N), Y = (y_1,
\dots ,y_N)$ in the single-spin configuration space, there exists a chain of
$m$ points $\{ Y = X_m, X_{m-1}, \dots , X_2, X_1 = X \}$ in the
configuration space, such that the product of matrix elements satisfies
$$\Pi^{X_m} K \Pi^{X_{m-1}} \dots \Pi^{X_3} K \Pi^{X_2} K \Pi^{X_1} \not=
0, \eqno(2.12)$$
with $X_1 \not= X_2 \not= X_3 \not= \dots \not= X_m$.}

{\bf Proof:}  We first consider a geometric question.  Place $N$ unlabelled
markers on the points $(x_1, \dots , x_N)$ of $\Lambda$.  The goal is to
move these markers, one at a time, across bonds of the lattice $\Lambda$ to
a final set of points $(y_1, \dots , y_N)$ in such a way that at no step is
there ever a doubly occupied site of the lattice.  We want to emphasize
that it is {\it not} necessary that the marker that was first at $x_1$ ends
up at $y_1$, we require {\it only} that in the final configuration the
sites $(y_1, \dots ,y_N)$ are occupied by some marker.

To do this we apply the following algorithm repeatedly --- at most $N$
times.  Look for the smallest $i$ such that the site $y_i$ is unoccupied.
Look for the smallest $j$ such the site $x_j$ is not in the set $\{ y_1, \dots
y_N \}$.  We will move a marker from the point $x_j$ and establish a marker
at the point $y_i$ in such a way that the other occupied sites of the final
state are identical to the other occupied sites of the initial state.
(Once again, we point out that the marker originally at $x_j$ need not end up
at $y_i$, and the other markers may be moved in this process.)  To achieve
this we choose a connected path $P$ in the lattice $\Lambda ,P = (x_j = z_1,
z_2, \dots z_k = y_i)$ from site $x_j$ to $y_i$.  Such a path exists by
hypothesis.  If there are no markers on the sites $z_2, \dots , z_{k-1}$,
then we simply move the marker at $x_j$ along the path $P$ to $y_i$.
Suppose on the contrary, that there are some other markers on this path
$P$.  Let $l$ be the largest number such that $z_l$ has a marker on it.
Then simply move this marker along the path to the site $y_i$, thereby
achieving two things:  a marker on $y_i$ and one less marker along the path
$P$.  We then move in turn each marker on the path $P$ to the location of
the previously moved marker.  This completes the description of the
algorithm and answers the geometric question.

To prove the lemma itself, we first note that $\langle \phi_2 \vert
\Pi^{X_1} K  \Pi^{X_2} \vert \phi_1 \rangle = 0$ for all $X_1$ and $X_2$
unless there is a permutation of $X_1$ such that after the permutation
there exists some $1 \leq j \leq N$ such that $x^1_i =
x^2_i$ for all $i \not= j$.  In this case
$$\langle \phi_2 \vert \Pi^{X_1} K \Pi^{X_2} \vert \phi_1 \rangle =
t_{ab} \phi_2 (X_1)^* \phi_1 (X_2) N! \eqno(2.13)$$
where $x^1_j = a$ and $x^2_j = b$.

The configurations $X_1, X_2, \dots X_m$ used in Eq. (2.12) will be the
$X$'s determined by the sequence of moves in the geometric discussion
above.  (Note that although the markers were indistinguishable there we
can, if we wish, put numbers on them.  In general the final state in this
case will not be identically the state $Y = (y_1, \dots y_N)$ but will be
some permutation $\widetilde Y$ of the final state.  On the other hand
$\Pi^Y = \Pi^{\widetilde Y}$ because the $L_x$'s commute.)  What remains to
be shown is that the operator
$$\Pi^{X_m} K \Pi^{X_{m-1}} \cdots \Pi^{X_2} K \Pi^{X_1} = (\Pi^{X_m} K
\Pi^{X_{m-1}}) (\Pi^{X_{m-1}} K \Pi^{X_{m-2}}) \cdots (\Pi^{X_2} K
\Pi^{X_1}) \eqno(2.14)$$
is nonvanishing.  Let $\phi_1, \dots, \phi_d$ be the orthonormal basis
proportional to antisymmetrized delta-functions in the configuration space.
{}From (2.13) we see that when we expand (2.14) in these intermediate states,
that a factor such as $\langle \phi_i \vert \Pi^{X_k} K \Pi^{X_{k-1}} \vert
\phi_j \rangle$ can be nonzero for only one $i$ and $j$, and
this contribution is equal to $t_{ab}$ (up to an overall sign) where $a$
and $b$ denote the two indices where $X_k$ and $X_{k-1}$ differ.  So, in
short, there will be exactly one matrix element of $\langle \phi_i \vert
\Pi^{X_m} K \Pi^{X_{m-1}} \dots \Pi^{X_2} K \Pi^{X_1} \vert \phi_j
\rangle$, and it will be a product of $t_{xy}$'s all of which are nonzero.
{\bf Q.E.D.}

We turn now to the proof of Theorem 2 for the Hubbard model.  The four
assumptions (i)-(iv) for the electron-phonon model simplify to the two
assumptions:  (a) $\Lambda$ is connected; (b) every $U_x$ is positive $(U_x
> 0)$.

{\bf Proof of Theorem 2 for the Hubbard model.}  Suppose $\Psi_1$ and
$\Psi_2$ are orthogonal ground states.  We can assume both $\Psi_1$ and
$\Psi_2$ are Hermitian.  Then $\Psi = \Psi_1 + \lambda \Psi_2$ is a ground
state for all real $\lambda$.  Moreover $\Psi$, if viewed as a matrix,
cannot be either positive or negative semidefinite for all real $\lambda$.
Hence for
some choice of $\lambda$ we have that the two ground states $\Psi_\pm :=
{1 \over 2} (\vert \Psi \vert \pm \Psi)$, are both nonzero, and are both ground
states, since $\vert \Psi \vert$ is a ground state from Theorem 1.  In
particular $\Psi_+$ satisfies the matrix Schr\"odinger equation
$$K \Psi_+ + \Psi_+K - \sum \limits^{\vert \Lambda \vert}_{x=1} U_x L_x
\Psi_+ L_x = e \Psi_+, \eqno(2.15)$$
and is a positive semidefinite matrix.

We define $\H_+$ to be the range of the matrix $\Psi_+$, which is a subspace
of $\H = \C^d$.  We also define $\H_\bot$ to be the orthogonal component of
$\H_+$ and $\Pi_\bot$ to be the projector onto $\H_\bot$.  By assumption
$\H_+$ and $\H_\bot$ are both nontrivial.

Multiply the Schr\"odinger equation (2.15) on the left and on
the right by $\Pi_\bot$ to yield $\sum \nolimits_x U_x \Pi_\bot L_x \Psi_+
L_x \Pi_\bot = 0$ since $\Pi_\bot \Psi_+ = \Psi_+ \Pi_\bot = 0$.  This
implies further that
$$\Psi_+ L_x \Pi_\bot = 0 \eqno(2.16)$$
since every $U_x$ is positive and $\Psi_+$ is a positive semidefinite
matrix.  Eq. (2.16) implies that the matrix $L_x$ does not connect the
subspaces $\H_+$ and $\H_\bot$.  Now multiply (2.15) on the right by
$\Pi_\bot$ and use (2.16) to discover that
$$\Psi_+ K \Pi_\bot = 0. \eqno(2.17)$$
Both (2.16) and (2.17) show that $\H_+$ and $\H_\bot$ are invariant subspaces
of the matrices $K$ and $L_x$.

We will now use the lemma to show that $\H_\bot$ is trivial, thereby
establishing a contradiction.  The operators $\Pi^X$ defined in (2.10) are
{\it one-dimensional} projectors in $\H$ and satisfy
$$(N!)^{-1} \sum \limits_X \Pi^X = \1. \eqno(2.18)$$
Thus $V^+ = \Pi^X \Psi^+$ is nonzero for some $X$.  Likewise, if $\phi \in
\H_\bot$ then $V^\bot = \Pi^Y \phi \not= 0$ for some $Y$.  By the lemma,
the rank one operator in (2.12), call it $A$, is nonzero and satisfies
$(V^\bot, A V^+) \not= 0$.  This is a contradiction since $A$, being a
product of $L_x$'s and $K$, has $\H_+$ as an invariant subspace. {\bf Q.E.D.}
\bigskip
\centerline{\bf III. ELECTRON-PHONON MODEL PROOFS}
\bigskip

The proof of Theorem 1 employs the same methods as utilized
for the Hubbard model in the preceding section, but now all of the expansion
coefficients and basis functions have an implicit ${\bf q}$ dependence.
The variational principle that shows that the ground state includes
a spin-singlet
state will arise from the kinetic energy terms for the phonons, since there is
no direct electron-electron interaction here (however, see the remark at
the end of the proof of Theorem 2 about including {\it attractive}
electron-electron interactions).

{\bf Proof of Theorem 1 for the electron-phonon model.}  Any wavefunction
$\Psi (\q)$ can be thought of as a matrix-valued function of $\q \in
\R^\nu$.  By $SU(2)$ invariance of (1.1) we can, as in the preceeding
proof, assume that there are equal numbers of up-spin and down-spin
electrons, namely $N$ of each kind.  The dimension $d$ of the matrix $\Psi
(\q)$ is then
$$d = {\vert \Lambda \vert \choose N}. \eqno(3.1)$$
As before, by taking transposes and complex conjugates, we can restrict our
discussion to the case where $\Psi (\q)$ is a Hermitian matrix for all
$\q$ (it is here that we use the condition that the hopping matrix elements
$t_{xy}$ are real because, without this condition, the complex conjugate of
$\Psi$ will generally have a different energy from that of $\Psi$). Note
that $\Psi (\q)$ is Hermitian, but it need not be real.

We can write the Schr\"odinger equation for $\Psi$ in the following generic
way
$$-\sum \limits^\nu_{j=1} {1 \over 2M_j} {\partial^2 \over \partial q^2_j}
\Psi (\q) + V(\q) \Psi (\q) + \Psi (\q) V(\q) = e \Psi (\q). \eqno(3.2)$$
This equation is to be understood in the following sense:  $\Psi (\q)$ is a
matrix-valued function and so are its second derivatives; $V(\q)$ is also a
matrix-valued function which is self-adjoint for every $\q$; the two terms
$V\Psi + \Psi V$ (with matrix multiplication being understood) include all
of the terms in $\H$ besides the phonon kinetic energy, $\sum
\limits^\nu_{j=1} p^2_j /2M_j$; of course $e$ is the energy
eigenvalue.

Associated with (3.2) is a variational expression which we can write as $\E
(\Psi) /\langle \Psi \vert \Psi \rangle$.  The denominator has the form
$$\langle \Psi \vert \Psi \rangle = \int_{\R^\nu} \Tr \Psi (\q)^2 \d \q.
\eqno(3.3)$$
The numerator is
$$\E (\Psi) = \int_{\R^\nu} \left\{ \sum \limits^\nu_{j=1} {1 \over 2M_j}
\Tr [\partial_j \Psi (\q)]^2 + 2 \Tr V(\q) \Psi (\q)^2 \right\} \d \q,
\eqno(3.4)$$
where $\partial_j$ denotes the partial derivative $\partial /\partial q_j$.

Our strategy is to replace the matrix $\Psi (\q)$, for every $\q$, by its
absolute value in the matrix sense, i.e.,
$$\vert \Psi (\q) \vert = \sqrt{\Psi (\q)^2}. \eqno(3.5)$$
We note that the norm satisfies $\langle \Psi \vert \Psi \rangle = \langle
\vert \Psi \vert \big\vert \vert \Psi \vert \rangle$, and that the $V
\Psi^2$ term is evidently unchanged.  In the appendix we prove that this
substitution does not increase the integral of $\Tr [\partial_j \Psi
(\q)]^2$.  There are some nontrivial technical issues here caused by the
fact that $\partial_j \Psi (\q)$ and $\partial_j \vert \Psi (\q) \vert$
may only be distributional derivatives, but these issues are fully resolved
in the appendix.

Since, by definition of the ground-state energy, $\E (\Psi)$ cannot
decrease when $\Psi$ is replaced by $\vert \Psi \vert$, we conclude that
$\vert \Psi \vert$ is also a ground state of the Hamiltonian.  As in the
Hubbard model proof, we conclude that $\vert \Psi (\q) \vert$ has a nonzero
projection onto the $S=0$ subspace, i.e., if every diagonal element of the
matrix $\vert \Psi (\q) \vert$ vanished for almost every value of $\q$,
$\vert \Psi (\q) \vert$ would be the zero function, which it is not. {\bf
Q.E.D.}

{\bf Proof of Theorem 2 for the electron-phonon model}.  Suppose there are
two ground states $\Psi_1 (\q)$ and $\Psi_2 (\q)$ which are Hermitian and
linearly independent.  Then $\Psi_1 + \lambda \Psi_2$ is a nonzero ground
state for every real $\lambda$, and as $\lambda$ is varied from $-\infty$ to
$+\infty$ there will be values of $\lambda$ for which $\Psi = \Psi_1 + \lambda
\Psi_2$ has both a negative and positive spectrum for a set of $\q$'s of
positive measure.  Then $\vert \Psi \vert$ is also a ground state, as are
$\Psi_+ := {1 \over 2} (\vert \Psi \vert + \Psi)$ and $\Psi_- := {1 \over 2}
(\vert \Psi \vert - \Psi)$.  Note that $\Psi, \vert \Psi \vert$, $\Psi_+$ and
$\Psi_-$ are all nonzero functions that satisfy the Schr\"odinger equation
(3.2).  Indeed $\Psi = \Psi_+ - \Psi_-$.  Moreover, both $\Psi_+ (\q)$ and
$\Psi_- (\q)$ are positive semidefinite matrices for all values of $\q$.
{}From the fact that the matrix valued function $V(\q)$ appearing in (3.2) is
differentiable, elliptic regularity theory applied to the Schr\"odinger
equation tells us that $\Psi (\q)$ is twice continuously differentiable
[actually only H\"older continuity of $V(\q)$ suffices for this
conclusion].

At each point $\q$, the vector space $\C^d$, on which $\Psi ({\bf q})$
operates,
is naturally the direct sum of three subspaces (some of which might be
empty).  These subspaces are denoted by $\H_+ (\q), \H_- (\q)$, and $\H_0
(\q)$.  $\H_+ (\q)$ is the spectral subspace of $\Psi_+ (\q)$, i.e.
consists of all linear combinations of the nonzero eigenvectors of $\Psi_+
(\q)$; $\H_- (\q)$ is the spectral subspace of $\Psi_- (\q)$; and $\H_0
(\q)$ is the orthogonal complement of $\H_+ (\q) \oplus\H_- (\q)$, i.e., the
subspace of all linear combinations of the zero eigenvectors of $\Psi_+
(\q)$ and $\Psi_- (\q)$.  Of course $\C^d = \H_+ (\q) \oplus \H_0 (\q) \oplus
\H_- (\q)$.

Let $d_0 (\q)$ denote the dimension of $\H_0 (\q)$ and let $d_0$ denote the
minimum of $\{ d_0 (\q): \q \in \R^\nu \}$.  Since there are only finitely
many values for $d_0 (\q)$ there is a point $\q_0 \in \R^\nu$ for which $d_0
(\q_0) = d_0$.  By definition the matrix
$\Psi (\q_0)$ has $d_0$ zero eigenvalues, and
the positive eigenvalues are separated from zero by a gap $\Delta$.  Since
$\Psi$, and hence the eigenvalues of $\Psi$, are continuous, there is some
ball $\B^\prime$ centered at $\q_0$ with radius $r$ such that the dimension
of $d_0 (\q) \leq d_0$ for every $\q \in \B^\prime$.  Since, however $d_0$
is the minimum of $d_0 (\q)$, we conclude that $d_0 (\q) = d_0$ for every
$\q \in \B^\prime$.

Now let us study the contour integral
$$\Pi_\bot (\q) := {1 \over 2 \pi i} \int_\Gamma {1 \over \Psi (\q) - z} \d
z \eqno(3.6)$$
where the contour $\Gamma$ runs from $-\infty$ just below the negative real
axis, goes vertically upward when the real part of $z = \Delta /2$ and
returns to $-\infty$ just above the negative real axis.

First we observe that $\Pi_\bot (\q_0)$ is the projector onto the orthogonal
complement of $\H_+ (\q_0)$, namely onto $\H_\bot (\q_0) = \H_0 (\q_0)
\oplus \H_-
(\q_0)$.  Furthermore, the zero eigenvalues of $\Psi_+ (\q)$ do not move from
zero as long as $\q \in \B^\prime$, as we have just seen above.  Therefore,
$\Pi_\bot (\q)$ continues to be the projector onto $\H_\bot (\q)$ as long as
the positive eigenvalues of $\Psi (\q)$ are greater than $\Delta /2$, and
hence do not intersect the contour $\Gamma$.  Since $\Psi$ is continuous we
conclude there is a smaller ball $\B \subset \B^\prime$ centered at
$\q$ in which
$\Pi_\bot (\q)$ continues to be the projector onto $\H_\bot (\q)$.  Since
$\Psi (\q)$ is twice continuously differentiable, it is a trivial matter to
show that we can differentiate under the integral sign in (3.6) and
conclude that $\Pi_\bot (\q)$ {\it is a twice continuously differentiable
matrix-valued function on} $\B$.

Now we compute some derivatives in this ball $\B$.  We start with the
observation that $\Psi_+ (\q) \Pi_\bot (\q) = 0$ for all $\q$ in $\R^\nu$.
The following identities hold in $\B$:  (we suppress the $\q$ dependence):
$$\eqalignno{(\partial_j \Psi_+) \Pi_\bot &+ \Psi_+ \partial_j \Pi_\bot = 0
\qquad&(3.7)\cr
(\partial^2_j \Psi_+) \Pi_\bot &+ 2 (\partial_j \Psi_+) \partial_j \Pi_\bot
+ \Psi_+ \partial^2_j \Pi_\bot = 0. \qquad&(3.8)\cr}$$
If the Schr\"odinger equation (3.2) for $\Psi_+ (\q)$ is multiplied on the
left and on the right by $\Pi_\bot (\q)$ we have (since $\Psi_+ \Pi_\bot =
0 = \Pi_\bot \Psi_+)$
$$\sum \limits^\nu_{j=1} {1 \over 2M_j} \Pi_\bot (\partial^2_j \Psi_+)
\Pi_\bot = 0, \eqno(3.9)$$
or, combining (3.9) with (3.8) we discover that
$$\sum \limits^\nu_{j=1} {1 \over 2M_j} \Pi_\bot (\partial_j \Psi_+)
\partial_j \Pi_\bot = 0. \eqno(3.10)$$
Now multiply (3.7) on the left by $\partial_j \Pi_\bot /2M_j$ and sum
over $j$.  The first term vanishes because of the adjoint of (3.10).  The
second yields
$$\sum \limits^\nu_{j=1} {1 \over 2M_j} \partial_j \Pi_\bot \Psi_+
\partial_j \Pi_\bot = 0. \eqno(3.11)$$
Since $\Psi_+ (\q)$ is a positive semidefinite matrix (and all $M_j <
\infty$), we conclude from (3.11) that
$$\Psi_+ \partial_j \Pi_\bot = 0, \qquad j = 1, \dots , \nu \eqno(3.12)$$
for all $\q \in \B$.  Eq. (3.12) states that the range of the derivatives
of $\Pi_\bot$ lies in $\H_\bot$, or
$$\partial_j \Pi_\bot = \Pi_\bot \partial_j \Pi_\bot, \eqno(3.13)$$
and therefore that each term in (3.7) separately vanishes, i.e.,
$$(\partial_j \Psi_+) \Pi_\bot = 0. \eqno(3.14)$$
Differentiating (3.14) yields
$$(\partial^2_j \Psi_+) \Pi_\bot + (\partial_j \Psi_+) \partial_j \Pi_\bot
= 0. \eqno(3.15)$$
But $(\partial_j \Psi_+) \partial_j \Pi_\bot = (\partial_j \Psi_+) \Pi_\bot
\partial_j \Pi_\bot = 0$ from (3.13) and (3.14), so we finally conclude
that the range of the second derivative of $\Psi_+$ lies in $\H_+$, or
$$(\partial^2_j \Psi_+ )\Pi_\bot = 0 \eqno(3.16)$$
for all $q \in \B$.

If we multiply the Schr\"odinger equation (3.2) for $\Psi_+$ on the right
by $\Pi_\bot$ and use the identity (3.16) we find
$$\Psi_+ (\q) \left[ K (\q) + \sum \limits^{\vert \Lambda \vert}_{x=1} L_x
G_x (\q) \right] \Pi_\bot (\q) = 0. \eqno(3.17)$$
This says that the matrix $W (\q) := K (\q) + \sum \nolimits^{\vert \Lambda
\vert}_{x=1} L_x G_x (\q)$ does not connect $\H_+$ with $\H_\bot$.
Differentiating (3.17) shows that
$$(\partial_j \Psi_+) W \Pi_\bot + \Psi_+ (\partial_j W) \Pi_\bot + \Psi_+
W \partial_j \Pi_\bot = 0. \eqno(3.18)$$
But we know that the range of the derivatives of $\Psi_+$ lies in $\H_+$ and
that of the derivatives of $\Pi_\bot$ lies in $\H_\bot$, so the first and
last terms of (3.18) vanish because $W$ does not connect $\H_+$ with
$\H_\bot$.  Since the hopping matrix elements do not have any $\q$
dependence by assumption and the derivative of the electron-phonon
couplings is a rank $\vert \Lambda \vert$ matrix by assumption, we find
$$\Psi_+ L_x \Pi_\bot = 0 \qquad x = 1, \dots , \vert \Lambda \vert
\eqno(3.19)$$
and, from (3.17),
$$\Psi_+ K \Pi_\bot = 0. \eqno(3.20)$$

These two identities imply that both $\H_+ (\q)$ and $\H_\bot (\q)$ are
invariant subspaces of the matrices $K$ and $L_x$ for all $\q \in \B$.
Exactly as in the Hubbard model proof, we conclude that for every $\q \in
\B$ one of the two alternatives holds:  either $\H_+ (\q) = \{ 0 \}$ or $\H_-
(\q) = \{ 0 \}$.  Since the functions $\Psi_+$ and $\Psi_-$ are continuous,
the set on which $\Psi_+$ is nonzero is open and the set on which $\Psi_-$
is nonzero is open.  Therefore $\B$ contains an open set in which either
$\Psi_+ (\q)$ is identically zero or $\Psi_- (\q)$ is identically zero.
However this cannot happen since $\Psi_+$ and $\Psi_-$ are eigenfunctions
with locally bounded potentials $V(\q)$ and therefore
satisfy a unique continuation
theorem$^{17}$.  That is to say if $\Psi_+$ vanishes in some open set, it
vanishes in all of $\R^\nu$ contrary to our original assumption that $\Psi_+$
is not identically zero.  {\bf Q.E.D.}

{\it Remarks.}---(1) Both Theorems 1 and 2 continue to be valid if
any {\it attractive} Hubbard model terms are added to the
Hamiltonian in (1.1).  In other words, the electron-phonon model
and Hubbard model results do not interfere with each other as long as they are
attractive.

(2) The SSH model is shown here to have at least one spin-singlet
state among its ground states.  The ground state is not shown to be unique
here.  The difficulty for the uniqueness proof enters in the derivative
of the matrix $W$ in Eq. (3.18).

\centerline {\bf IV. DISCUSSION}

The results presented in this contribution hold only for an even number of
electrons in a finite system.  In this case, the only nonanalyticities that can
enter in properties of the ground state occur when there is a ground-state
level crossing.  In the cases where the ground state can be shown to be unique,
there are never any level crossings, so that any ``transition'' of the
electron-phonon ground state from a collection of delocalized polarons to
a collection of self-trapped polarons is not a sharp transition, but is
a smooth crossover.
We are not prepared to prove any statements about the thermodynamic
limit here.

One can ask if these results will survive if a magnetic field is turned on.
The answer in general is no because the variational argument presented
in Theorem 1 no longer holds if there are interactions with the
local electronic spin [as one would have if one added a Zeeman coupling
to the Hamiltonian in Eq. (1.1)].  One can also investigate the effect of
a magnetic field on the electronic kinetic energy.
The hopping matrix elements are always assumed to be real, and therefore
can only incorporate flux phases that correspond to entrapped fluxes that
are integral multiples of $\pi$.  If the hopping matrix elements become
complex, the entire methodology incorporated here fails, and there are no
rigorous statements that we can make about this case.

In conclusion, we have presented a proof that the ground state of a general
class of electron-phonon Hamiltonians must include a state that is a spin
singlet.  The phonons can interact with the electrons in two different
ways---the
phonons interact with the local electronic charge and the phonons modulate
the electronic hopping integrals.  The phonon coordinate dependence of the
both electron-phonon couplings and the hopping matrix elements is arbitrary.
The phonons can be optical modes or acoustical modes, and can contain
anharmonic
couplings.  The hopping matrix is always assumed to be real and symmetric.
The Su-Schrieffer-Heeger model, the Holstein model, and the Hubbard
model all fall into this general class.  In the case where the hopping matrix
contains no phonon coordinate dependence, the lattice is connected,
the electron-phonon couplings are independent, and the inverse phonon masses
are all positive, the ground state has also been shown to be unique.
\bigskip
\centerline {\bf Acknowledgments}
\bigskip

We are indebted to Jan Philip Solovej for valuable discussions, especially
for his help with the appendix.
This work was supported by the U. S. National Science Foundation under
grant no. PHY90-19433 A03 [EHL], and by the Office of Naval Research under
grant no. N00014-93-1-0495 [JKF].
\vfill\eject
\noindent
{\bf APPENDIX:  ABSOLUTE VALUE DECREASES KINETIC ENERGY}
\bigskip

We shall prove here that replacing a matrix-valued function,
$\Psi (\q)$ by its absolute value, $\vert \Psi (\q) \vert =
\sqrt{\Psi^\dagger (\q) \Psi (\q)}$, decreases each component of the kinetic
energy
$$T_j (\Psi) := \int \limits_{\R^\nu} \Tr [\partial_j \Psi^\dagger (\q)
\partial_j \Psi (\q)] \d \q, \eqno(A.1)$$
when $\Psi$ is Hermitian.  (Here $\partial_j = \partial /\partial q_j$ and
$\dagger$ denotes adjoint.)

Before going into the technicalities, let us give a heuristic discussion to
motivate the truth of our assertion.  If we write, in Dirac notation,
$$\Psi (\q) = \sum \limits^d_{\alpha =1} w_\alpha (\q) \vert f_\alpha (\q)
\rangle \langle f_\alpha (\q) \vert, \eqno(A.2)$$
where $d$ is the dimension of the matrix (in our proof later we shall
generalize to $d = \infty)$, the functions $w_\alpha$ are the real
eigenvalues and the $\vert f_\alpha (\q) \rangle$ are the $\q$-dependent
orthonormal eigenfunctions of the Hermitian matrix $\Psi (\q)$.  Supposing
everything to be differentiable we can compute
$$\partial_j \Psi (\q) = \sum \limits^d_{\alpha = 1} \partial_j w_\alpha
(\q) \vert f_\alpha (\q) \rangle \langle f_\alpha (\q)\vert + w_\alpha (\q)
\vert f_\alpha (\q) \rangle \langle g_\alpha (\q) \vert + w_\alpha (\q)
\vert g_\alpha (\q) \rangle \langle f_\alpha (\q) \vert, \eqno(A.3)$$
where $\vert g_\alpha (\q) \rangle = \partial_j \vert f_\alpha (\q)
\rangle$.  Since $\langle f_\alpha (\q) \vert f_\beta (\q) \rangle =
\delta_{\alpha \beta}$, we have that $\langle f_\alpha (\q) \vert g_\beta
(\q) \rangle + \langle g_\alpha (\q) \vert f_\beta (\q) \rangle = 0$.
Thus, if we square (A.3) and take the trace we have
$$\Tr [\partial_j \Psi (\q)]^2 = \sum \limits^d_{\alpha =1} [\partial_j
w_\alpha (\q) ]^2 + 2 w_\alpha^2 (\q) \langle g_\alpha (\q) \vert g_\alpha
(q) \rangle - 2 \sum \limits^d_{\alpha =1} \sum \limits^d_{\beta = 1}
w_\alpha (\q) w_\beta (\q) \vert \langle g_\alpha (\q) \vert f_\beta (\q)
\rangle \vert^2. \eqno(A.4)$$
{}From this we see that replacing $w_\alpha (\q)$ by $\vert w_\alpha (\q)
\vert$ can only decrease the last term on the right.  The second term does
not change.  The first term also does not change since, for any real,
differentiable function $w(\q)$, it is a fact that $[\partial_j w(\q)]^2 =
[\partial_j \vert w(\q) \vert ]^2$ in the sense of distributions.

Notice that this heuristic discussion gives a pointwise inequality $\Tr
[\partial_j \Psi (\q)]^2 \geq \Tr [\partial_j \vert \Psi (\q)\vert]^2$.  In
our rigorous discussion we shall content ourselves with an inequality for
the integral (A.1) --- which is sufficient for our purposes in this paper.

We begin the rigorous discussion with some definitions.  Let $\H$ be a
fixed, separable Hilbert space (for the purposes of our paper $\H$ is
finite dimensional, but there is no need for this restriction here).  Let
$\B$ denote the Hilbert-Schmidt operators on $\H$ and, for $A \in \B$, let
$$\Vert A \Vert := \{ \Tr A^\dagger A \}^{1/2}$$
denote its {\it Hilbert-Schmidt norm}.
We also define
$$\vert A \vert := \sqrt{A^\dagger A} \quad {\sl and} \quad \vert A^\dagger
\vert := \sqrt{AA^\dagger}$$
and note that $\Vert \vert A \vert \Vert = \Vert \vert A^\dagger \vert
\Vert = \Vert A \Vert$.

A map
$$\Psi : \R^\nu \rightarrow \B$$
is said to be {\it measurable} if the function $f_A (\q) := \Vert \Psi (\q)
- A \Vert$ is (Lebesgue) measurable for every Hilbert-Schmidt $A$.  It is
not hard to prove that $\Psi$ is measurable if and only if every matrix
element $(v, \Psi (\q) w)$ is a measurable function for every fixed $v$ and
$w$ in $\H$.  $\Psi$ is said to be in $L^2 (\R^\nu; \B)$ if $\Psi$ is
measurable and if $f_0 \in L^2 (\R^\nu)$, i.e., $\int f_0 (\q)^2 < \infty$.

The map $\Psi$ is said to be in $H^1 (\R^\nu ;\B)$ if $\Psi \in L^2
(\R^\nu; \B)$ and if there are maps $\partial_1
\Psi, \partial_2 \Psi, \dots , \partial_\nu \Psi$ for which the following holds
\item{(i)}  Each $\partial_j \Psi : \R^\nu \rightarrow \B$ is a map in $L^2
(\R^\nu; \B)$
\item{(ii)}  For each infinitely differentiable map, $\phi : \R^\nu
\rightarrow \B$, of compact
support with derivatives $\partial_j \phi$, we have the relation
$$\int \limits_{\R^\nu} \Tr [\partial_j \phi (\q) \Psi (\q)] \d \q = - \int
\limits_{\R^\nu} \Tr [\phi (\q) \partial_j \Psi (\q)] \d \q \eqno(A.5)$$
\smallskip\noindent
for each $j$.  (Note:  to say that $\phi$ is differentiable means that for
every $\q \in
\R^\nu$, $\lim \nolimits_{\varepsilon \rightarrow 0} \Vert \varepsilon^{-1}
[\phi (\q + \varepsilon \e_j) - \phi (\q)] - \partial_j \phi (\q) \Vert =
0$ with $\e_j$ being the unit vector in the $j^{th}$ direction.  Since
$\phi$ has compact support this limit is uniform in $\q$.)  Clearly
$\partial_j \Psi^\dagger = (\partial_j \Psi)^\dagger$.

It is easy to verify that $H^1 (\R^\nu; \B)$ is a
Hilbert space with inner product
$$(\Psi, \Psi^\prime) = \int \limits_{\R^\nu} \Tr [\Psi^\dagger (\q)
\Psi^\prime (\q) + \sum \limits^\nu_{j=1} \partial_j \Psi^\dagger (\q)
\partial_j \Psi^\prime (\q)] \d\q .\eqno(A.6)$$
Clearly, $\Psi$ is in $H^1$ (or in $L^2$) if and only if $\Psi^\dagger$ is
in $H^1$ (or in $L^2$).
This class, $H^1 (\R^\nu; \B)$, is precisely the class needed for quantum
mechanics, i.e., so that the variational energy $T_j (\Psi)$ can be defined
and so that the norm $\int \Tr \Psi^\dagger \Psi$ can be defined.  For
matrix--valued functions (i.e., $\H$ is finite dimensional) the properties
of measurability, being in $L^2$ and being in $H^1$ are
just the ordinary meaning of these properties applied to each matrix
element of $\Psi$ considered as a function on $\R^\nu$.

We are indebted to Jan Philip Solovej for very considerable help with the
following.

{\bf THEOREM.}  {\it Let $\Psi$ be in $H^1 (\R^\nu; \B)$.  Then $\vert \Psi
\vert$ and $\vert \Psi^\dagger \vert$ are in $H^1 (\R^\nu; \B)$ and, for
each $j$
$$2 T_j (\Psi) \geq T_j (\vert \Psi \vert) + T_j (\vert \Psi^\dagger
\vert). \eqno(A.7)$$
In particular, if $\Psi (\q)$ is self-adjoint for all $\q$ then
$$T_j (\Psi) \geq T_j (\vert \Psi \vert). \eqno(A.8)$$}

{\it Remark:}  If $\Psi$ is not self-adjoint it is quite possible that $T_j
(\Psi) < T_j (\vert \Psi \vert)$.  But then $T_j (\Psi) > T_j (\vert
\Psi^\dagger \vert)$.

{\it Proof:}  It suffices to prove the theorem when $\Psi (\q)$ is
self-adjoint for all $\q$.  To see this, consider the Hilbert space $\H_2 =
\H \oplus \H$ and replace $\Psi$ by the self adjoint operator
$\Psi_2 = \pmatrix{0&\Psi^\dagger \cr \Psi &0 \cr}$.  Then $\vert \Psi_2
\vert = \pmatrix{\vert \Psi \vert &0 \cr 0& \vert \Psi^\dagger \vert \cr}$
and (A.7) becomes $T_j (\Psi_2) \geq T_j (\vert \Psi_2 \vert)$.

The first task is to show that $\vert \Psi \vert$ is measurable.  Our
definition of measurability given above is the standard one, namely the
inverse image of an open ball in the Banach space $\B$ is measurable.  Now
the map $\Psi \rightarrow \vert \Psi \vert$ is continuous in the $\B$ norm
and hence, by a standard result, $\vert \Psi \vert$ is measurable.

Let $v_1, v_2 \dots$, be an orthonormal basis for $\H$.  Matrix elements
$(v_\alpha, A v_\beta)$ will be denoted by $A_{\alpha \beta}$.  It is easy
to verify that $\partial_j (\phi_{\alpha \beta}) = (\partial_j
\phi)_{\alpha \beta}$ in the classical sense if $\phi$ is
differentiable.  If $\Psi$ has distributional derivatives then $\partial_j
(\Psi_{\alpha \beta}) = (\partial_j \Psi)_{\alpha \beta}$ in the
{\it ordinary} distributional sense.  [Simply
take $\phi (\q) = h (\q) \vert v_\beta \rangle \langle v_\alpha \vert$ in
(A.5), with $h$ being an ordinary infinitely differentiable function of
compact support.]  Another important preliminary remark is that by Fubini's
theorem $\sum \limits_{\alpha, \beta} \int_{\R^\nu} K_{\alpha \beta} (\q)
\d \q = \int_{\R^\nu} \sum \limits_{\alpha, \beta} K_{\alpha \beta}
(\q) \d \q$ for any nonnegative, measurable
functions $K_{\alpha \beta} (\q)$.

For any $\Psi$ in $H^1 (\R^\nu ; \B)$ (including $\phi$, as a special case)
we define $\partial_{j \varepsilon} \Psi (\q) =
\varepsilon^{-1} [\Psi (\q + \varepsilon \e_j) - \Psi (\q)]$.
By the fundamental theorem of calculus for distributions
$$[\partial_{j \varepsilon} \Psi (\q)]_{\alpha \beta} = \partial_{j
\varepsilon} [\Psi (\q)_{\alpha \beta}] = \int^1_0 \partial_j [\Psi (\q + t
\varepsilon \e_j)_{\alpha \beta}] \d t = \int^1_0 [\partial_j
\Psi (\q + t \varepsilon \e_j)]_{\alpha \beta} \d t \eqno(A.9)$$
Thus, by Fubini's theorem and Schwarz's inequality,
$$\eqalignno{\int_{\R^\nu} \Tr &[\partial_{j \varepsilon} \Psi (\q)]^2 \d
\q = \sum \limits_{\alpha, \beta} \int_{\R^\nu} \left\{ \int^1_0
[\partial_j \Psi (\q + t \varepsilon \e_j)]_{\alpha \beta} \d t \right\}^2
\d \q \cr
&\leq \sum \limits_{\alpha, \beta} \int_{\R^\nu} \int^1_0 [\partial_j \Psi
(\q + t \varepsilon \e_j)]^2_{\alpha \beta} \d t \d \q = \int_{\R^\nu} \Tr
[\partial_j \Psi (\q)]^2 \d \q. \qquad&(A.10)\cr}$$
Therefore, $\partial_{j \varepsilon} \Psi$ is uniformly (in $\varepsilon$)
bounded in $L^2 (\R^\nu ; \B)$.

Now we are ready to study $\vert \Psi \vert$, and we begin with a little
lemma.  If $N$ and $M$ are self-adjoint Hilbert-Schmidt operators then
$$\Tr (N-M)^2 \geq \Tr (\vert N \vert - \vert M \vert)^2. \eqno(A.11)$$
This is equivalent to $\Tr NM \leq \Tr \vert N \vert \vert M \vert$.  If we
write $2 N_\pm = \vert N \vert \pm N$ and $2 M_\pm = \vert M \vert \pm M$,
we have that $N_\pm$ and $M_\pm$ are positive semidefinite, and our
requirement now reads $\Tr [M_+ N_- + M_- N_+] \geq 0$.  But this is true
because $\Tr M_+ N_- = \Tr M^{1/2}_+ N_- M^{1/2}_+ > 0$, etc.

{F}rom (A.11) we have that $\Tr [\partial_{j \varepsilon} \vert \Psi \vert]^2
\leq \Tr [\partial_{j \varepsilon} \Psi]^2$, from which we deduce that
$\partial_{j \varepsilon} \vert \Psi \vert$ is uniformly bounded in $L^2
(\R^\nu ; \B)$.
More precisely, by (A.10) and (A.11),
$$\int_{\R^\nu} [\partial_{j \varepsilon} \vert \Psi \vert (\q)]^2 \d \q
\leq \int_{\R^\nu} \Tr [\partial_j \Psi (\q)]^2 \d \q. \eqno(A.12)$$
Since $L^2 (\R^\nu ; \B)$ is a Hilbert space, the boundedness shown above
plus the Banach-Alaoglu theorem implies that there is a sequence
$\varepsilon_1, \varepsilon_2, \dots$, tending to zero such that
$$\partial_{j \varepsilon_n} \vert \Psi \vert \rightharpoonup \uprho_j,
\eqno(A.13)$$
as $n \rightarrow \infty$, where $\uprho_j$ is a map in
$L^2 (\R^\nu ;\B)$ and where the convergence is in the weak sense.
Since, $\partial_{j \varepsilon} \phi (\q)$ converges to $\partial_j \phi
(\q)$ uniformly (in Hilbert-Schmidt norm)
we can write (A.5), using (A.13), as
$$\int_{\R^\nu} \Tr [\phi \uprho_j] = \lim \limits_{\varepsilon
\rightarrow 0} \int_{\R^\nu} \Tr [\phi \partial_{j \varepsilon} \vert \Psi
\vert]
= -\lim \limits_{\varepsilon \rightarrow 0} \int_{\R^\nu} \Tr [\partial_{j
\varepsilon} \phi \vert \Psi \vert] = - \int_{\R^\nu} \Tr [\partial_j \phi
\vert \Psi \vert] \eqno(A.14)$$
The first equality in (A.14) is the weak convergence of $\partial_{j
\varepsilon} \Psi$; the second is just a trivial change of variables in the
$q$-integration; the third is the uniform convergence of $\partial_{j
\varepsilon} \phi$.  Equation (A.14) holds for every $\phi$.  By uniqueness
of the
distributional derivative for ordinary functions [and choosing $\phi (\q) =
h (\q) \vert v_\beta \rangle \langle v_\alpha \vert$ as before] we conclude
that $(\uprho_j)_{\alpha \beta} = (\partial_j \vert \Psi \vert)_{\alpha
\beta}$, and hence $\uprho_j = \partial_j \vert \Psi \vert$.  However,
norms cannot increase under weak limits and thus (A.8) follows from (A.12).
{\bf Q.E.D.}

\vfill\eject
\bigskip
\centerline {\bf References}
\bigskip
\item{$^1$} A. B. Migdal, Zh. Eksp. Teor. Fiz. {\bf 34,} 1438 (1958) [Sov.
Phys.---JETP {\bf 7,} 999 (1958)].
\item{$^2$} G. M. Eliashberg, Zh. Eksp. Teor. Fiz. {\bf 38,} 966 (1960)
[Sov. Phys.---JETP {\bf 11,} 696 (1960)].
\item{$^3$} W. L. McMillan and J. M. Rowell, Ch. 11, {\it Superconductivity}
, edited by R. Parks (Marcel Dekker, Inc., New York, 1969).
\item{$^4$} P. B. Allen and B. Mitrovi\' c, Solid State Phys. {\bf 37,}
1 (1982).
\item{$^5$} J. P. Carbotte, Rev. Mod. Phys. {\bf 62,} 1027 (1990).
\item{$^6$} T. Holstein, Ann. Phys. {\bf 8,} 325 (1959).
\item{$^7$} W. P. Su, J. R. Schrieffer, and A. J. Heeger, Phys. Rev. B
{\bf 22}, 2099 (1980).
\item{$^8$} E. H. Lieb, Phys. Rev. Lett. {\bf 62,} 1201 (1989).
\item{$^9$} E. H. Lieb and D. C. Mattis, Phys. Rev. {\bf 125,} 164 (1962).
\item{$^{10}$} H. L\" owen, Phys. Rev. B {\bf 37,} 8661 (1988).
\item{$^{11}$} L. M. Falicov and J. C. Kimball, Phys. Rev. Lett. {\bf 22,}
997 (1969).
\item{$^{12}$} T. Kennedy and E. H. Lieb, Physica {\bf 138A,} 320 (1986);
E. H. Lieb, Physica {\bf 140A,} 240 (1986).
\item{$^{13}$}  E.H. Lieb and M. Loss, Duke Math. J. {\bf 71}, 337 (1993).
\item{$^{14}$}  J.L Lebowitz and N. Macris, {\it Low temperature phases of
itinerant fermions interacting with classical phonons:  the static Holstein
model}, J. Stat. Phys. {\bf 76} to appear (1994).
\item{$^{15}$} S. Aubry, G. Abramovici, and J.-L. Raimbault, J. Stat. Phys.
{\bf 67,} 675 (1992).
\item{$^{16}$} J. Hubbard, Proc. Royal Soc. London (Ser. A) {\bf 276,} 238
(1963).
\item{$^{17}$}  M. Reed and B. Simon, {\it Methods of Modern Mathematical
Physics, Vol. IV:  Analysis of Operators}, (Academic Press, New York, 1978)
p. 226.
\vfill
\bye